\documentclass[a4paper]{jpconf}
\usepackage{graphicx, epsf, epsfig, amssymb}
\usepackage[usenames,dvipsnames]{xcolor}
\usepackage{color}
\usepackage{aas_macros}
\usepackage[breaklinks]{hyperref}
\usepackage{amsfonts,amsmath,amssymb,mathrsfs}
\def\be{\begin{equation}}
\def\ee{\end{equation}}
\def\beq{\begin{eqnarray}}
\def\eeq{\end{eqnarray}}

\newcommand{\msun}{M_\odot}

\newcommand\gsim{\mathrel{\rlap{\raise 0.511ex \hbox{$>$}}{\lower 0.511ex\hbox{$\sim$}}}}

\def\nn{\nonumber}

\begin{document}
\title{Massive Black Hole Science with eLISA}

\author{Enrico Barausse$^{1,2}$, Jillian Bellovary$^{3,4}$, Emanuele Berti$^{5}$, Kelly Holley-Bockelmann$^{3,4}$, Brian Farris$^{6,7}$, Bangalore Sathyaprakash$^{8}$, Alberto Sesana$^{9}$}

\address{$^{1}$CNRS, UMR 7095, Institut d'Astrophysique de Paris, 98bis Bd Arago, 75014 Paris, France}
\address{$^{2}$Sorbonne Universit\'es, UPMC Univ Paris 06, UMR 7095, 98bis Bd Arago, 75014 Paris, France}

\address{$^{3}$Department of Physics and Astronomy, Vanderbilt University, Nashville, TN 37235, USA}
\address{$^{4}$Department of Physics, Fisk University, Nashville, TN 37208, USA}

\address{$^{5}$Department of Physics and Astronomy, The University of Mississippi, University, MS 38677, USA}

\address{$^{6}$Center for Cosmology and Particle Physics, Physics Department, New York University, New York, NY 10003, USA}
\address{$^{7}$Department of Astronomy, Columbia University, 550 West 120th Street, New York, NY 10027, USA}

\address{$^{8}$School of Physics and Astronomy, Cardiff University, 5, The Parade, Cardiff, CF24 3AA, United Kingdom\footnote{Currently on sabbatical leave at LIGO Laboratory, California Institute of Technology, MS 100-36, Pasadena, CA 91125}}

\address{$^{9}$Max-Planck-Institut f\"ur Gravitationsphysik, Albert Einstein Institut, D-14476, Golm, Germany}

\begin{abstract}
The evolving Laser Interferometer Space Antenna (eLISA) will
revolutionize our understanding of the formation and evolution of
massive black holes (MBHs) along cosmic history, by probing massive
black hole binaries (MBHBs) in the $10^3-10^7\msun$ range out to
redshift $z\gtrsim 10$ . High signal-to-noise ratio detections of
$\sim 10-100$ MBHB coalescences per year will allow accurate
measurements of the parameters of individual MBHBs (such as their
masses, spins and luminosity distance), and a deep understanding of
the underlying cosmic MBH parent population. This wealth of
unprecedented information can lead to breakthroughs in many areas of
physics, including astrophysics, cosmology and fundamental physics. We
review the current status of the field, recent progress and future
challenges.

\end{abstract}

\section{Introduction}



The evolving Laser Interferometer Space Antenna (eLISA, \cite{lisa13})
is designed to be sensitive to gravitational waves (GWs) at mHz
frequencies. One of the strongest sources in this frequency window are
MBHBs merging throughout the Universe
\cite{2013GWN.....6....4A}. According to our current understanding of
structure formation in a $\Lambda$CDM Universe, MBHBs frequently form
along cosmic history following galaxy mergers. MBHs we see in today's
galaxies are expected to be the natural end-product of a complex
evolutionary path, in which black holes (BHs) seeded in proto-galaxies
at high redshift grow through cosmic history via a sequence of MBHB
mergers and accretion episodes
\cite{2000MNRAS.311..576K,2003ApJ...582..559V}. However, our current
observational knowledge of the MBH population is limited to a small
fraction of these objects: either those that are active (see
e.g. \cite{2010A&A...518A..10V}), or those in our neighborhood, where
stellar- and gas-dynamical measurements are possible (see
\cite{2013ARA&A..51..511K} for a review).

eLISA will revolutionize this picture by probing MBHBs in the
$10^3-10^7\msun$ range out to redshift $z\gtrsim 10$
\cite{2013GWN.....6....4A}. In the current design \cite{lisa13}, eLISA
will be capable of detecting $\sim 10-100$ MBHB coalescences per year
and of accurately measuring the parameters of individual MBHBs, such
as their masses, spins and luminosity distance. This wealth of
unprecedented measurements has the potential to revolutionize many
areas of physics, ranging from astrophysics to cosmology and
fundamental physics. This paper summarizes contributions to the ``LISA
Symposium X'' that were devoted to this subject.

We start in Section \ref{sec:kelly-jillian} by describing current
models for MBH formation and evolution, highlighting present
uncertainties due either to the lack of observations or to poor
theoretical constraints. eLISA will dramatically change this situation
by probing the very first coalescences of seed BHs at high
redshift. The interaction of MBHs with the environment has a key role
in bringing MBHBs close enough that GW emission becomes
efficient. Section~\ref{sec:brian} describes recent developments on
this front, focusing on observational signatures of merging MBHBs in
the electromagnetic (EM) domain. Coincident detection of MBHB mergers
as both GW {\em and} EM sources will pave the way to multimessenger
astronomy, promising extraordinary advances in the understanding of
accretion physics. In Section~\ref{sec:enrico} we show how eLISA will
open a new era of precision measurements of MBH spins. Individual spin
measurements will allow us to test gravity in the strong-field regime
with unprecedented accuracy; the {\em collective} properties of spin
and mass distributions of the whole MBH population carry information
on the physics of accretion flows, and in general of the intimate link
between MBHs and galaxy evolution across cosmic
time. Section~\ref{sec:sathya} explores eLISA's potential for
cosmology. GW observations can yield a direct measurement of the
luminosity distance to the source. If an EM counterpart is detected,
the coincident measurement of the source redshift will provide an
appealing opportunity for calibration-free cosmography. Last but not
least, Section~\ref{sec:emanuele} touches on the invaluable insights
that eLISA will provide in terms of fundamental physics. Dynamical
measurements of the behavior of gravity in the strong-field regime
will constrain the geometry of the spacetime around BHs,
telling us whether the Kerr solution actually describes these objects
and potentially even yielding smoking guns of new physics beyond
General Relativity (GR).

%
%
%

\section{Birth and growth of massive black holes}
\label{sec:kelly-jillian}

There is mounting observational evidence that supermassive BHs
with masses between $10^6$ and $10^{10} M_{\odot}$ reside at the heart
of nearly every galaxy
\cite{Gehren84,kormendyho13,ferrarese05}. Though MBHs are an
observational certainty, nearly every aspect of their evolution --
from their birth, to their fuel source, to their basic dynamics -- is
a matter of lively debate.

The existence of bright quasars at high redshift implies that billion
solar mass BHs must be in place within their host galaxies
less than a billion years after the Big Bang. This remarkable
observational fact suggests that the seeds of these most massive MBHs
are sown during, or even before, the formation of
protogalaxies. However, the precise MBH seed formation mechanism is
not known, nor is it clear that there is only one seed formation
channel at play over the entire MBH mass spectrum.

Currently, the two most widely accepted MBH birth scenarios are a) the
remnants of the first generation of stars (Population III: see
e.g.~\cite{Bond84, Couchman86,mr01,Abel02,Bromm04}), and b) the direct
collapse of pristine gas within massive dark matter
halos~\cite{Loeb94,Oh2002,Begelman2006,Lodato06,Wise2008,Choi2013}. In
both these models, overdensities within a gas cloud collapse as the
gas radiates away energy; if the gas can cool efficiently, then it
gravitationally fragments into ever smaller clouds. The mass of the
final collapsed object, be it a protostellar cloud or a seed black
hole, is determined by the Jeans mass of the gas that can no longer
cool efficiently enough to fragment further.

In the Population III seed scenario, metal-free gas is cooled by H$_2$
and HD, but since these molecules are much less effective than metals
in radiating away energy, the Jeans mass is likely to be larger than
that of stars formed in the present era. However, one of the
difficulties in this scenario is that the actual initial mass function
of this first generation of stars is hotly debated, and the issue will
not easily resolve itself with the current state-of-the-art
simulations. Early work placed the stellar mass at $100 - 1000 \,
M_\odot$~\cite{Abel2002,Bromm2002}, but later work suggested that
fragmentation is far more
common~\cite{Stacy2010,Jappsen2009,Machida2008,Greif12,Susa14}. A
recent study examining the role of turbulence in Population III star
formation suggested a flat initial mass function with a characteristic
mass less than $100 \, M_\odot$ (e.g.~\cite{Clark2011}). The newest
work tends to indicate that, even though the protostellar clouds do
fragment, they tend to quickly merge within a dynamical time,
essentially generating the very massive Population III stars once
again~\cite{Inayoshi14,InayoshiOmukaiTasker14}. While these
theoretical advances are promising, a more thorough treatment
involving radiative transfer and 3D magnetohydrodynamics is required
before we will truly understand the nature of the first stars and
whether they can spawn MBH seeds.


In the fiducial direct collapse model, the only coolant is atomic
Hydrogen, and this requires that the gas sit in dark matter halos with
virial temperatures higher than $10^4$ K. To inhibit fragmentation and
to suppress H$_2$ formation, there is an additional photon bath that
heats the gas; one of the most likely candidates for this radiation
field, the Lyman-Werner background, is thought to be generated by
nearby Population III stars~\cite{Dijkstra2008, Regan09, Shang10} at
redshifts z $\sim 10-20$.  Preventing efficient cooling and
fragmentation in this way results in a huge Jeans mass, and the
eventual seed BHs are of order $10^4-10^6 \, M_\odot$. These
direct collapse BHs are the most obvious candidates to grow
into the MBHs powering high-redshift quasars, because they require no
stringent assumptions on the accretion or merger history to grow to
$10^9 \, M_\odot$ by $z \sim 6$.

As with the Population III scenario, the main epoch of direct collapse
seed formation is brief, ending as metals pollute the halo gas and the
radiation field drops below a critical heating threshold as the
Universe expands \cite{Yue2014,Ferrara2014}. Still, pockets of
pristine and irradiated gas could remain at low redshift that could
collapse into MBH seeds~\cite{Jimenez06,Bellovary11}.
One potential issue with this direct collapse model is that the
intensity of radiation needed to prevent fragmentation is under
debate, and the actual Lyman-Warner background generated at redshift
$z\sim 20$ is not well-constrained, as it depends in part on the
occupation fraction and initial mass function of Population III
stars. Therefore, direct collapse BH formation may arguably be
too rare to account for the quasar population, much less the entire
MBH mass spectrum (see however \cite{Visbal2014}).

One of the great promises of a future space-based GW observatory like
eLISA is that it may be able to pin down the relative efficiency of
light versus MBH seed formation
channels~\cite{Schneider2000,Sesana07,2011PhRvD..83d4036S,2011MNRAS.415..333P}.
These pathways affect the occupation fraction of MBHs in local dwarf
galaxies, the existence of intermediate-mass BHs, and the
scaling relations between MBHs and their host galaxies (i.e. the
$M$--$\sigma$ relation) at low masses. Studies have attempted to
predict MBH-galaxy occupation fractions \cite{Tanaka09,Bellovary11}
and the low-mass $M$--$\sigma$ relation \cite{Volonteri09} for a range
of seed models, but observations in this regime are very difficult,
and thus constraints are difficult to obtain. Direct GW detections
will elucidate much regarding the hidden MBH population and their
seeding mechanism.

Turning now to MBH growth, it is tempting to think that a clearer
picture emerges. The famous Soltan argument has often been invoked to
claim that MBHs are fueled nearly entirely by gas during the brief
quasar epoch. The logic goes like this. If we assume that quasars are
powered by gas accretion onto MBHs, then we can turn the observed
energy and number density of optically-bright quasars into an estimate
of the gas mass accreted by MBHs during the quasar era. Happily, if we
compare the mass density of accreted gas during the quasar phase
(assuming a radiative efficiency of 10\%) to the mass density locked
up in the local MBH population, the numbers agree. This implies that
90\% of MBH masses are built from gas that is accreted at the
Eddington limit before redshift 2; it also implies that MBHs do not
seem to accrete in a low-efficiency, or ``quiet'' mode, nor are there
many obscured or undetected quasars within the MBH mass budget. Though
this argument seems pat and iron-clad, there are many uncertainties
involved. For example, converting the optically observed quasar
luminosity function to a bolometric energy density is fraught with
difficulty. Furthermore, it is very clear that optical quasar surveys
do miss a large fraction of the real quasar population, those that are
radio-loud or X-ray bright (see e.g.~\cite{Brusa10,Eckart10}).  In
addition, many theoretical efforts call into question the assumption
of Eddington-limited accretion, some advocating super-Eddington
accretion along filaments to feed ultramassive BHs
\cite{Priya09}, and others promoting a ``radio'' mode of quiescent gas
accretion \cite{Hardcastle07,Merloni08}.  Along this line, simulations
indicate that the high-redshift universe may be rife with hidden MBHs
that are accreting at less than the Eddington
rate~\cite{Bellovary13,Micic2011}.

Despite these uncertainties, it is widely accepted that gas is the
primary fuel for MBH growth. It is thought that galaxy mergers help to
restock the gas reservoir around the BH as the gas is shocked
and then falls toward the galactic center. Indeed one of the best
explanations of the M--$\sigma$ relation invokes galaxy mergers that
drive gas toward the BH, which fuels it and subsequently
generates a prodigious radiative ``feedback'' that pushes the
remaining gas toward the galaxy outskirts and cuts off the MBH fuel
source (see e.g.~\cite{Hopkins06}). Once the gas is in the galaxy
bulge potential, it slowly cools and forms new stars with high
velocity dispersion \cite{DiMatteo05,Mayer07}. Though early
simulations of equal-mass galaxy mergers were very promising in
supporting this view, later unequal-mass mergers fail to grow the
primary MBHs enough to fall on the M-$\sigma$ relation
\cite{VanWassenhove12,Micic2011}. The problem is that most low-mass
MBHs, like the one in our Milky Way, simply don't undergo equal-mass
mergers at the rate needed to build the MBH and to couple this growth
to the bulge in this framework.

Fear not: mergers are not the only process which may trigger MBH
growth.  In fact, galaxies which lack a bulge (often thought to have
extremely quiescent merger histories) may host MBHs as well, and even
actively growing ones.  One thought is that MBHs may grow efficiently
through the accretion of cold, unshocked gas (i.e. ``cold flows'')
which enter the galaxy through filamentary accretion.  Some galaxies
acquire most of their gas through this process
\cite{Keres05,Brooks09}, and MBHs may efficiently accrete this gas as
well \cite{DiMatteo12,Bellovary13}.  Another method which may be
prevalent consists of mergers of MBHs with other MBHs \cite{KHB2010}.
In the simulations presented in \cite{Bellovary13}, the central MBHs
in massive galaxies at $z = 4$ have built up over half of their mass
through MBHB mergers.  This is a regime where GW observations can
verify the accuracy of these predictions, which depend on many
properties of MBH seed formation (initial mass, formation redshift,
and efficiency of formation).  By directly collecting GW data from the
growth of high-redshift MBHs, we can determine if this phenomenon is
dominated by MBH mergers, or whether it is correlated with galaxy
merger rates, or some combination of the two.

Overall, it is likely that there are several mechanisms to grow a MBH,
depending on the assembly history of the host galaxy.  The most
massive MBHs are likely fueled via gas accretion from major mergers at
high redshift \cite{Treister12}, which explains the relative tightness
of the M--$\sigma$ relation in this regime.  Lower-mass galaxies,
which experience fewer major mergers, exhibit more scatter within the
local scaling relations, likely because their MBHs grow through more
stochastic processes such as cold flow accretion, or through secular
processes like bar-driven gas inflow. The combination of these
processes plus the unknown contribution of MBHB mergers leads to a
myriad of possibilities for MBH growth, which will be possible to
disentangle by means of low-frequency GW observations.

\section{Gas accretion onto massive black hole binaries and their electromagnetic signature}
\label{sec:brian}

According to the evolution picture emerging from Section
\ref{sec:kelly-jillian}, MBHB coalescences must be common events at
all redshifts. These systems do not live in isolation, but they are
embedded in dense galactic nuclei, surrounded by gas and
stars. Therefore, the interaction of MBHBs with their environment may
provide a unique opportunity to observe EM signatures as well as GWs,
opening new avenues in multimessenger astronomy. Information from a
simultaneous detection of EM and GWs may be useful for studying
fundamental aspects of gravitational physics.  For example, in some
modified gravity scenarios, the propagation velocity for gravitons may
differ from that of photons \cite{kocsis08,deffayet07}. Additionally,
the measurement of the luminosity distance from the GW signal at an
accuracy of $1-10\%$, coupled with the redshift information from the
EM detection, could serve as a cosmological ``standard siren'' of
unprecedented accuracy (better than $\sim 1 \%$) \cite{holz05}.  Such
detections may also combine accurate measurements of MBH spins and
masses obtained from GW signals with EM observations to probe MBH
accretion physics in great detail \cite{kocsis06}. Since most eLISA
sources will be relatively low-mass systems at high redshift, where
gas-rich environments are common, we focus here on the interaction
between a MBH and a putative massive circumbinary disk.

The standard picture of circumbinary accretion disks can be described
as follows. Tidal torques from the binary tend to drive gas outward,
clearing an evacuated cavity in the innermost region of the
disk. Meanwhile, viscous torques transport angular momentum outward in
the disk, allowing gas to flow inward and refill this cavity. The
balance of tidal and viscous torques determines the location of the
inner edge of the circumbinary disk at $r\approx 2 a$, where $a$ is
the binary separation. This balance can be maintained, provided the
timescale $t_{\rm gw}$ for inspiral of the binary due to GW emission
is much longer than the viscous timescale of the disk, $t_{\rm
  visc}$. This is known as the ``pre-decoupling'' epoch.

To date, a number of ``dual'' systems in which two MBHs occupy the same
galaxy but are too widely separated to be gravitationally bound have
been observed \cite{komossa03, comerford13,liu13,woo14}, as well as
several candidate binary systems (see
e.g. \cite{2012AdAst2012E...3D,liu14} and references
therein). Proposed EM signatures of such binaries include spatially
resolving two AGN-like point sources, identifying double-peaked broad
emission lines, spatial structures in radio jets, characteristic time
variability in quasar emission, and characteristic features in quasar
spectra.

Theoretical aspects of this problem have been studied analytically
\cite{goldreich80,artymowicz94,armitage02,milos05,chang10,haiman09,shapiro10,kocsis12a,tanaka13},
often using approximate angle-averaged tidal torque formulae. While
these techniques have proven very useful in highlighting qualitative
features of the accretion, they tend to overestimate the barrier to
accretion imposed by binary torques by imposing symmetry in the
accretion flow. As EM counterparts to MBH mergers depend sensitively
on the amount of gas available for accretion, it is important to
understand non-axisymmetric effects which may provide a mechanism for
delivering more gas to the MBHs. Indeed, 2D and 3D simulations have
demonstrated that gas streams may be stripped from the inner cavity,
accreting directly onto the binary. Prior numerical simulations have
been performed in 2D \cite{macfadyen08,dorazio13} and in 3D
\cite{hayasaki07,cuadra09,farris11,farris12,noble12,shi12,roedig12}. While
3D codes have been useful in probing the gas dynamics during the final
orbits prior to MBH merger, long viscous timescales render them
prohibitively costly for simulating the quasi-steady-state flow during
the ``pre-decoupling'' epoch. For this epoch, 2D simulations such as
those performed using the {\it DISCO} code \cite{duffell14} have
proven quite useful. These simulations include the inner cavity in the
computational domain, and use shock-capturing Godunov-type methods to
evolve thin ($h/r \sim 0.03$) disks over the viscous timescales
necessary to accurately capture the steady-state accretion at high
resolution.

The {\it DISCO} code is a moving-mesh code which allows one the
freedom to specify the motion of the computational cells
\cite{duffell13}. For binary accretion, the chosen rotation profile
matches the nearly Keplerian fluid motion outside the cavity, while
transitioning to uniform rotation at the binary orbital frequency
inside the cavity. Because grid cells move azimuthally with the fluid,
advection errors are minimized, allowing for the accurate capture of
the dynamics of accretion streams which penetrate into the cavity.

The quasi-steady-state solutions which the simulations relax to after
several viscous timescales can be interpreted as generalizations of
the Shakura-Sunyaev disk solutions \cite{shakura73}, with the central
gravitating object replaced by a binary. The initial disk
configurations consist of the ``middle region'' Shakura-Sunyaev
solution for a steady-state, geometrically thin, optically thick
accretion disk, assuming a gas-pressure dominated fluid with electron
scattering as the dominant opacity. The fluid evolves according the
the 2D viscous Navier-Stokes equations, assuming an $\alpha$-law
viscosity prescription. A $\Gamma$-law equation of state of the form
$P=(\Gamma - 1)\epsilon$ is chosen, where $\epsilon$ is the internal
energy density, and the adiabatic index is set to
$\Gamma=5/3$. Appropriate radiative cooling and viscous heating terms
are added to the energy equation. The cooling rate for an optically
thick, geometrically thin disk is $q_{cool} = 4 \sigma/3\tau T^4$,
where $T$ is the mid-plane temperature, and $\tau$ is the optical
depth for electron scattering ($\tau=\Sigma\sigma_T/m_p$). It is
assumed that the fluid is gas-pressure dominated everywhere, ignoring
radiation pressure. In each simulation the binary is chosen to have
zero eccentricity.

\begin{figure}[h]
    \includegraphics[width=18pc]{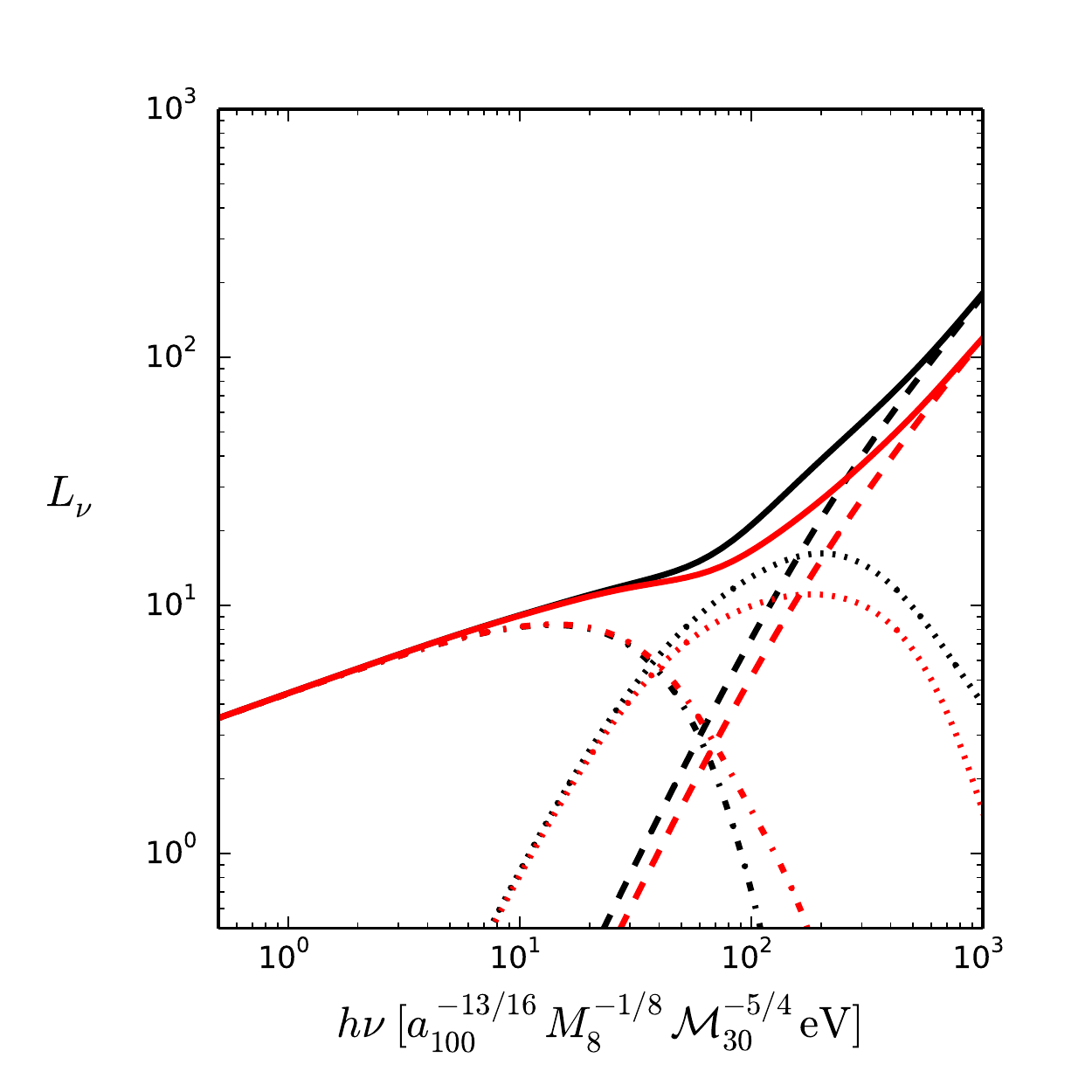}\hspace{2pc}%
    \begin{minipage}[b]{16pc}\caption{\label{fig:farris_spectrum}Thermal spectra computed from simulation snapshot at $t \gtrsim t_{vis}$. Red curves are calculated from disk data $\approx 3 t_{bin}$ after that of black curves. The full spectrum is represented by solid lines, the component arising only from minidisk regions within distance $d<0.5a$ of either MBH is represented by dashed lines, the ``cavity'' emission is represented by dotted lines, and the emission from the ``outer region'' is represented by dashed-dotted lines. We have introduced the scaling parameters $a_{100}\equiv a/100M$, $M_8 \equiv M / 10^8 M_{\odot}$, and $\mathcal{M}_{30} \equiv \mathcal{M}_a / 30$.}
    \end{minipage}
\end{figure}

The important findings of these simulations are the following:
\begin{enumerate}
    \item As shown in Fig.~\ref{fig:farris_density}, gas enters the
      circumbinary cavity along accretion streams, and the interaction
      of these streams with the cavity wall causes the cavity to
      become lopsided (in agreement with
      e.g.~\cite{macfadyen08,cuadra09,roedig12,shi12,noble12,farris14}). See
      \cite{farris14} and \cite{shi12} for a description of the
      mechanism driving the growth of this lopsidedness.
\begin{figure}[h]
    \includegraphics[width=18pc]{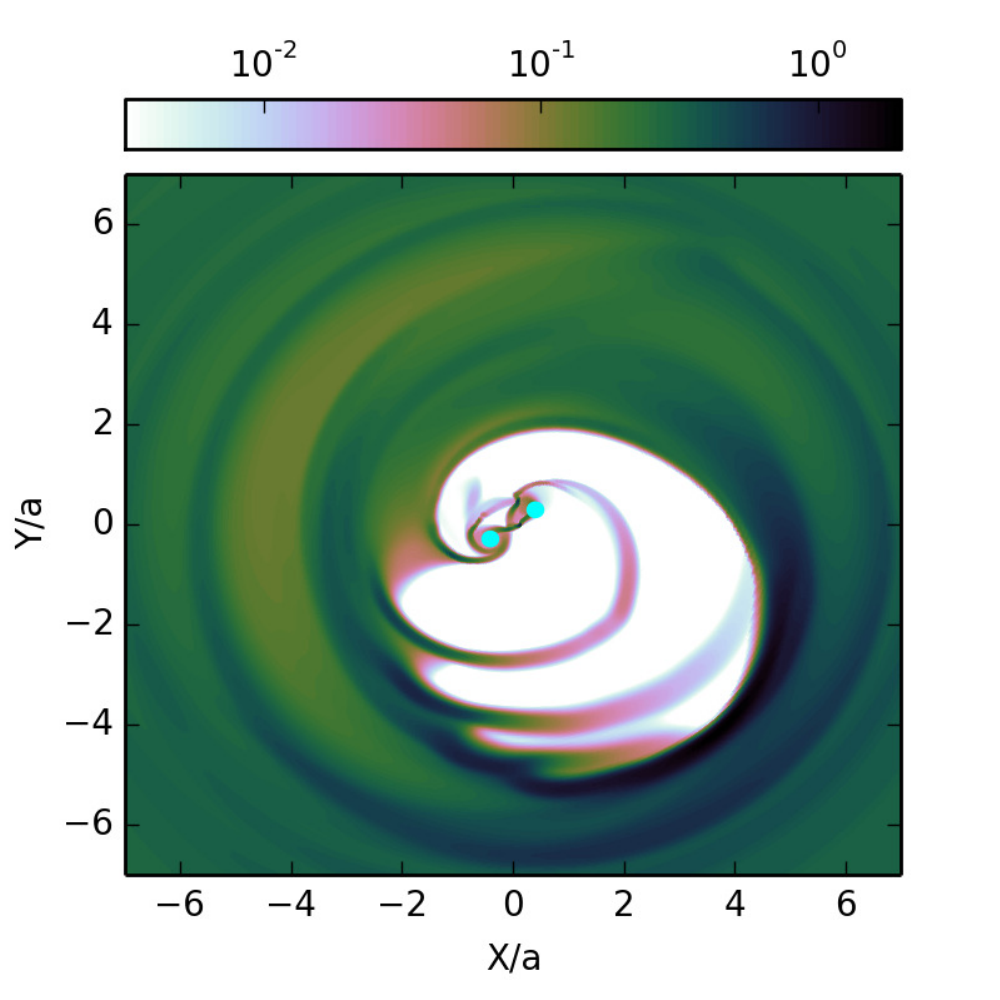}\hspace{2pc}%
    \begin{minipage}[b]{16pc}\caption{\label{fig:farris_density}Snapshot of surface density $\Sigma$ during quasi-steady state after $t \gtrsim t_{vis}$. Surface density is normalized by the maximum value at $t=0$ and plotted on a logarithmic scale in the inner $\pm 6a$. Orbital motion is in the counter-clockwise direction.}
    \end{minipage}
\end{figure}
    \item These simulations support the growing consensus that the
      accretion rate onto the MBHs is not significantly reduced by the
      presence of a binary, when compared to the accretion rate onto a
      single MBH of the same mass
      \cite{dorazio13,shi12,noble12,roedig12}. This is the case in
      spite of the fact that much of the inner cavity is cleared of
      gas by the action of the binary torques, and it is due to the
      effectiveness of the narrow accretion stream in delivering gas
      from the circumbinary disk inner edge to the individual black
      holes.
    \item For each mass ratio considered, ``mini-disks'' surrounding
      each MBH are formed. In each case, the mini-disks are
      persistent, as their accretion timescale greatly exceeds the
      binary orbital timescale. For the binary mass ratios $q=0.11$
      and $q=0.43$, the size of these mini-disks is in rough agreement
      with the semi-analytic predictions of \cite{artymowicz94}.
    \item Significant periodicity in the accretion rates emerges for
      $q \gtrsim 0.1$. At these mass ratios, the binary torques are
      strong enough to excite eccentricity in the inner cavity and
      create an overdense lump, whose interaction with the passing
      MBHs leads to periodicity in the accretion rate.  The strongest
      peak in the periodograms for these cases corresponds to the
      orbital frequency of the lump, with many associated harmonics
      for the $q\gtrsim 0.43$ cases. This periodicity may constitute a
      unique observational signature of MBHBs.
    \item For each case considered, the accretion rate onto the
      secondary is sufficiently large relative to that of the primary,
      so that the mass ratio $q$ is increasing. Similar results have
      been found previously in SPH calculations
      \cite{hayasaki07,cuadra09,roedig11,roedig12}. As MBHs are
      expected to gain a significant fraction of their mass through
      gas accretion \cite{soltan82,yu02,elvis02}, this suggests a
      mechanism which may bias the distribution of binaries near
      merger toward higher mass ratios.
    \item The emission associated with the shock heating in the
      accretion streams is sufficient to bring the emission from
      within the cavity above that of a disk around a single MBH. The
      peak in $dL/dr$ which appears at $r/a \approx 5$ corresponds to
      the shock heating of gas in the stream which is not directly
      accreted, but rather impacts the cavity wall as seen along the
      lower-right edge of the cavity in Fig.~\ref{fig:farris_density},
      leading to the thin strip of bright emission. Scaled to a $10^8
      M_{\odot}$ binary with a separation near decoupling at $a/M =
      100$, this enhancement is significant in soft and hard X-rays
      (see Fig.~\ref{fig:farris_spectrum}). This X-ray enhancement has
      been predicted by \cite{roedig14}, who estimated the
      characteristic frequency of mini-disk ``hot spot'' emission by
      estimating the amount of energy released due to shock heating
      when accretion streams impact the minidisks. Instruments
      sensitive to X-ray emission from AGN such as
      XMM-Newton\footnote{http://sci.esa.int/xmm-newton/},
      NuSTAR\footnote{http://www.nustar.caltech.edu/}, the upcoming
      eROSITA\footnote{http://www.mpe.mpg.de/erosita/} all-sky survey,
      and the proposed
      ATHENA\footnote{http://www.the-athena-x-ray-observatory.eu/}
      X-ray observatory may be sensitive to these signatures.
\end{enumerate}

While there remain no confirmed observations of EM counterparts for
MBHBs, simulations have already provided tantalizing evidence that
significant gas can accrete onto such a binary, even at separations in
the eLISA band. A number of distinguishing observational signatures of
such systems have been proposed, including directly resolving two
AGN-like point sources, searching for offset emission lines, spatial
structures in radio jets, characteristic time variability in quasar
emission, and characteristic features in quasar spectra. As we enter
the age of GW astronomy the possibility of a simultaneous measurement
of both gravitational and EM radiation from a merging MBHB appears
increasingly likely.

\section{Massive black hole spins as gravitational and cosmological probes}
\label{sec:enrico}

In GR, MBHs are described by the Kerr metric and are completely
characterized by three ``charges'' (or ``hairs''), the mass $M$, the
spin $S$ and an electric charge $Q$, with the spin satisfying the Kerr
bound $a=cS/(G M^2)\leq1$ (if the spin is larger than this bound, the
Kerr metric does not describe a BH but a naked singularity). Also, the
charge $Q$ is usually expected to be negligible for astrophysical BHs,
due to the presence of plasma that quickly neutralizes it, and also
because of quantum effects such as Schwinger pair production
or vacuum  breakdown mechanisms 
producing cascades of electron-positron pairs near the BH horizon. BHs
in gravity theories different from GR are still characterized by the
mass and spin, but may also present additional charges that might
provide a way to test gravity in strong-field regimes (cf.~Section
\ref{sec:emanuele} for a related discussion).

The effect of the mass in GR is qualitatively similar to Newtonian
gravity, i.e. it sources an attractive force decaying as $GM/r^2$ for
$r\gg G M/c^2$ ($r$ being the distance from the BH). Because of the
long-range character of the force it sources, the mass can be
estimated rather easily with EM observations, e.g. (in the case of
MBHs) by observations of the nuclear dynamics of stars and gas,
fitting of the spectral energy distribution of galaxies, spectroscopic
single epoch measurements and reverberation mapping.
EM measurements of MBH masses are typically accurate only within
$20-50\%$ (e.g. even for our own SgrA$^{*}$, the mass can only be
estimated to within $\sim 10$\%),
but still allowed discovering correlations with galactic properties,
hinting at a symbiotic co-evolution of MBHs with their galactic
hosts~\cite{gebhardt00, ferrarese00, marconi03, haring04, gultekin09}.

Measuring BH spins is more complicated, because they do not enter the
dynamics at Newtonian order, but only at 1.5 post-Newtonian order,
i.e.~they change the Newtonian equations of motion only by corrections $\sim
{\cal O}(v/c)^3$, where $v$ is the characteristic velocity of the gas
and/or stars surrounding the MBH. This results in significant effects
only very close to the event horizon. The most promising EM technique
to estimate the spins of MBHs is through the spectra of
relativistically broadened K$\alpha$ iron lines
(cf.~\cite{reynolds13,brenneman13} for recent reviews on this
topic). Indeed, X-ray observatories such as XMM-Newton and Suzaku have
by now measured significant samples of spins, and more constraints are
becoming available from NuSTAR's hard X-ray
data~\cite{risaliti13,marinucci14a,marinucci14b}. The masses and spins
of MBHs in binary systems will be measured very accurately by eLISA,
with errors $\delta M\lesssim 0.1\%$ and $\delta a\lesssim
0.01$~\cite{2009CQGra..26i4027A,2013GWN.....6....4A}. These GW
measurements are very clean compared to EM ones, because they are not
affected by the systematics usually present in EM data (which are due
to poor understanding of the dissipative gas physics). This is because
GWs are emitted in the latest stages of a binary system's evolution,
when the effect of the gas on the orbital dynamics is typically
negligible~\cite{2014PhRvD..89j4059B,2014arXiv1404.7140B}.

More specifically, eLISA will detect GWs from MBHBs both during the
long post-Newtonian inspiral, which radiates in the eLISA band for
separations less than about 200 gravitational radii for binaries with
total mass $\sim 10^5-10^6 M_\odot$ (i.e. those to which eLISA is most
sensitive), and during the final plunge, merger and ringdown of the
system. This will allow testing a plethora of strong-field
general-relativistic effects, namely:
\begin{enumerate}
    \item The spin-orbit coupling (also known as ``frame
      dragging'')~\cite{1975PhRvD..12..329B,1994PhRvD..49.6274A},
      which causes the BH spins to precess during the inspiral, thus
      producing amplitude modulations in the emitted gravitational
      waveforms~\cite{1994PhRvD..49.6274A,2004PhRvD..70d2001V,2006PhRvD..74l2001L,2011PhRvD..84b2002L,2008ApJ...677.1184L},
      which increase the signal-to-noise ratio and improve the
      estimation of the source parameters (and in particular the sky
      localization). The precessional dynamics can be very rich: for
      example, to a first approximation (and on short timescales) the
      spins precess around the binary's total angular momentum, but in
      certain configurations they can undergo a more complicated
      ``transitional precession''~\cite{1994PhRvD..49.6274A} (whereby
      spins almost anti-aligned with the orbital angular momentum
      change their orientation on a very short timescale at some
      point during the inspiral), or get locked in secularly stable
      resonant configurations that tend to align or anti-align the
      spins~\cite{2004PhRvD..70l4020S,2010PhRvD..81h4054K,2010ApJ...715.1006K,2012PhRvD..85l4049B};
    \item The spin-spin
      coupling~\cite{1995PhRvD..52..821K,1993PhRvD..47.4183K}, which
      appears at higher post-Newtonian order than the spin-orbit
      coupling and thus modifies the spin precession in the later
      inspiral, causing additional modulations in the gravitational
      waveforms and improving parameter estimation;
    \item The non-linear relations between the final mass, spin and
      recoil velocity of the BH remnant forming from the merger, and
      the masses and spin vectors of the binary's components at large
      separations. These relations have been studied in detail with
      fully general-relativistic numerical simulations (see
      e.g.~\cite{2010RvMP...82.3069C,2011arXiv1107.2819S,2014arXiv1405.4840L}
      for reviews on this topic), whose results can be extrapolated to
      generic binary configurations by exploiting knowledge of the test-particle,
      self-force and post-Newtonian
      dynamics~\cite{2008PhRvD..77b6004B,2008PhRvD..78h4030K,2008PhRvD..78d4002R,2009ApJ...704L..40B,2008PhRvD..78h1501T,2014arXiv1406.7295H,2012ApJ...758...63B,2010ApJ...719.1427V};
    \item The quasi-normal mode ringing of the BH remnant (see
      ~\cite{2009CQGra..26p3001B} for a review), whose frequencies and
      decay times are functions of the remnant's mass and spin alone
      if GR is correct. Measuring these frequencies therefore
      constitutes a genuine strong-field test of the gravity theory
      (cf. Section~\ref{sec:emanuele}).
\end{enumerate}

Besides testing the strong-field general-relativistic dynamics through
these effects, eLISA measurements of the spins will also provide
useful information on the cosmological evolution of MBHs. In fact,
although the impact of gas on the dynamics of MBHBs is negligible once
they enter the eLISA
band~\cite{2014PhRvD..89j4059B,2014arXiv1404.7140B}, radiatively
efficient accretion of gas is known to be the main driver of the
evolution of the masses and spins on cosmological
timescales~\cite{soltan82, merloni04, shankar04, shankar13b}. Also,
because eLISA will detect GWs from MBHBs at redshift $z\gtrsim 10$,
mass and spin measurements will give precious information about the
high-redshift formation of the first generation of BH seeds, from
which present-day MBHs are believed to descend.

Early studies~\cite{2011PhRvD..83d4036S} in this direction highlighted
for instance that eLISA mass measurements will be able to discriminate
with high confidence between a scenario in which MBHs evolve from
``heavy'' ($\sim 10^4-10^5 M_\odot$) seeds forming at $z=10-15$ from
the collapse of protogalactic disks, and a ``light-seed'' scenario in
which the seeds form at $z\sim 20$ from the collapse of Population III
stars into BHs with masses $\sim 50-300 M_\odot$, shedding light on
the nature of the first seed BHs forming in the young universe
(cf. Section \ref{sec:kelly-jillian}). As for the spin evolution,~\cite{2008ApJ...684..822B} (see also~\cite{volonteri05,lagos2009,
  fanidakis11, fanidakis2}) showed that eLISA should be able to tell a
coherent accretion scenario (where gas accretion always happens on
prograde orbits on the equatorial plane) from a chaotic scenario
(where the MBH captures clouds that are isotropically distributed
around it with randomly oriented angular momenta). Clearly, the first
scenario predicts almost extremal spins (because accretion is always
prograde), while the second predicts small spins $a\lesssim 0.3$,
because the angular momentum transferred by the clouds tends to cancel
out on long timescales.

Two ingredients were missing from these early attempts, namely
\textit{(a)} the strong-field spin-orbit coupling reviewed above, and
\textit{(b)} a more realistic connection between the properties of the
accretion flow and those of the host galaxy. Regarding \textit{(a)},~\cite{bardeen75,king05,king06,perego09} showed that the
interaction between the spin-orbit coupling and the viscous stresses
active inside an off-equatorial, geometrically thin accretion disk
results in a quick alignment between the MBH spin $\boldsymbol{S}$ and
the disk's orbital angular momentum $\boldsymbol{L}$ (a phenomenon
known as ``Bardeen-Petterson effect''), provided that
$|\boldsymbol{L}|> 2 |\boldsymbol{S}|$. Because this alignment takes
place on a timescale much shorter than the accretion timescale,
accretion will be essentially coherent if a MBH is hosted in a
gas-rich galactic nucleus (where it is more likely that the condition
$|\boldsymbol{L}|> 2 |\boldsymbol{S}|$ will be satisfied), while in
gas-poor nuclei accretion will be more likely to resemble the chaotic
accretion scenario outlined above. The Bardeen-Petterson effect is
also expected to be important for the orientation of the spins of the
merging binaries detectable with eLISA. Indeed, binaries forming in
gas-rich nuclei will likely have almost aligned spins, and thus
produce lower recoil velocities for the merger remnant, which is
therefore unlikely to escape from the host
galaxy~\cite{2007ApJ...661L.147B}.

An investigation of the impact of effects \textit{(a)} on the cosmological
evolution of the MBH spins was first performed by~\cite{2012MNRAS.423.2533B}, which also accounted for
\textit{(b)} by simulating the co-evolution between the MBHs and their
host galaxies with a semi-analytical galaxy formation model, including
both the evolution of dark-matter (via merger trees) and the evolution
of baryonic structures (intergalactic and interstellar media, galactic
disks, star-forming spheroids, as well as MBHs with their accretion
disks). Ref.~\cite{2012MNRAS.423.2533B} also obtained predictions (in
principle testable with eLISA) for the fraction of MBH mergers in
gas-rich environments (and thus with aligned spins) as opposed to ones
in gas-poor environments (and thus with misaligned spins). However, in
spite of its sophistication, the model of~\cite{2012MNRAS.423.2533B}
still assumed that the accretion flow onto the MBH had vanishing
average angular momentum in gas-poor regimes (i.e. whenever the
condition $|\boldsymbol{L}|> 2 |\boldsymbol{S}|$ is not
satisfied). This assumption is an idealization, as galaxies do have a
non-zero angular momentum, and one would therefore expect the average
angular momentum $\bar{L}$ of the clouds accreting onto the MBH to be
non-zero. Allowing for $\bar{L}\neq0$ can indeed have important
consequences for the spin evolution of MBHs, as shown in~\cite{dotti13}.

Ref.~\cite{dotti13} left $\bar{L}$ as a free parameter, because a
first-principle calculation would be extremely challenging: galaxy
formation simulations are not yet capable of resolving the MBH's
sphere of influence and the accretion disk, and in any case it is far
from clear whether such simulations have full control of the
subgrid/dissipative physics at small scales (see
however~\cite{dubois13} for a recent attempt to extract the angular
momentum of the clouds from a hydrodynamical simulation, with
resolution up to $10$ pc.) Alternatively, a bold attempt can be made
at trying to connect $\bar{L}$ to measurements of the velocity
dispersion $v/\sigma$ of the gas and stars in galaxies (e.g.~if
$v/\sigma$ were zero, it is clear that accretion would be perfectly
isotropic, i.e. $\bar{L}=0$). Ref.~\cite{SBDR} adopted this approach
and used the semi-analytical galaxy formation model
of~\cite{2012MNRAS.423.2533B} with the spin-evolution model
of~\cite{dotti13}, connecting $\bar{L}$ to measurements of $v/\sigma$
of the gas and stellar components in various galaxy
morphologies. While still debatable because these measurements are
currently only available at distances of $\gtrsim 100$ pc from the MBH
(and thus far from the accretion disk and the MBH's sphere of
influence), this approach allowed the authors of~\cite{SBDR} to
produce testable predictions for the spin distribution in competing
models for the ``isotropy'' $\bar{L}$ of the accretion flow. In
particular, three models were considered: $(A)$ one connecting
$\bar{L}$ to the velocity dispersion of the gaseous component of
galaxies; $(B)$ one connecting $\bar{L}$ to the velocity dispersion of
the stellar component; and $(C)$ a hybrid model. A comparison with
existing iron-K$\alpha$ measurements of MBH spins shows that model
$(A)$ is ruled out quite convincingly, while both models $(B)$ and
$(C)$ are in agreement with observations (with marginally significant
evidence in favor of the hybrid model $(C)$). Ref.~\cite{SBDR} also
showed that the idealized accretion prescriptions discussed above
(namely purely coherent and purely chaotic accretion, as well as the
original model of~\cite{2012MNRAS.423.2533B}) are disfavored by
existing iron-K$\alpha$ measurements. Clearly, eLISA's measurements of
the spins of merging binaries will allow discriminating between these
competing models with much higher significance, thus providing a way
to test the properties of the accretion flow onto MBHs with
unprecedented accuracy.

\section{Cosmography with space-based detectors}
\label{sec:sathya}

The geometry, large-scale structure and dynamics of the Universe can
be inferred with precision if we can accurately measure the distance
and redshift to sources distributed throughout the Universe.  This is
because the luminosity distance $D_{\rm L}$ to a source as a function
of its redshift $z$ depends on the geometry of the Universe and a
number of cosmological parameters, such as the Hubble parameter,
relative fractions of density in dark energy, dark matter, baryonic
matter, etcetera~\cite{Liddle:2003}. Redshift is very well measured by
spectroscopic methods, except for sources at low redshifts ($z\ll 1$),
where peculiar velocities due to the gradient of the local
gravitational potential could be comparable to cosmological expansion.
The real challenge, however, is to accurately measure distances to
cosmological sources.

Precision cosmography is enabled by sources whose intrinsic luminosity
$L$ can be deduced by some observed property of a source, e.g. its
time variability, so that one can infer the luminosity distance
$D_{\rm L}$ from its apparent luminosity $F,$ namely $D_{\rm L} =
\sqrt{L/4\pi F}.$ In 1986 Schutz pointed out how GW observations of
inspiralling compact binaries could provide an astronomer's ideal
standard candle \cite{Schutz:1986}. Since then there has been quite a
lot of work in trying to understand how observations of mergers
involving MBHs by LISA (and eLISA) could be used to measure
cosmological parameters
\cite{Holz:2005df,Dalal:2006ab,Arun07b,Petiteau:2011we}. Here we will
briefly review the basic idea, challenges posed by observations and
some solutions. The current perspective on the problem is that eLISA's
application for cosmography will be limited by weak gravitational
lensing, and precision measurements of cosmological parameters are
only possible if eLISA observes several tens of sources during its
lifetime. In this regard, ground-based detectors are likely to be more
useful for cosmology as they are expected to observe a very large
number of sources, which helps to mitigate problems posed by weak
lensing \cite{Sathyaprakash:2009xt}.

\subsection {Self-calibrating standard sirens}
\label{sec:cosmography:basic idea}
Inspiralling compact binaries are often referred to as
self-calibrating standard sirens \cite{holz05}. The word ``siren'' is
used to indicate that GWs are more akin to sound waves than EM waves,
and eLISA is a detector that ``listens'' to its sources. They are
``self-calibrating'' as no other calibration process is required to
measure the distance to an inspiralling binary other than the
description of its dynamics by GR \cite{Schutz:1986}. This favorable
situation should be contrasted with the astrophysical modelling of
sources and construction of a cosmic distance ladder, that are
required to calibrate the distance to cosmological sources
\cite{Liddle:2003}.

The response of eLISA to GWs from the inspiral phase of a coalescing
binary depends on its distance $D_{\rm L}$ from the Earth, sky
position $(\theta, \varphi),$ component masses $(m_1, m_2)$ and their
spins, orbital eccentricity and orbital angular momentum $(\vec S_1,
\vec S_2, e, \vec L)$ (all at some fiducial time) \cite{holz05}. In
order to measure the luminosity distance to the source it is necessary
to disentangle all the other parameters. The shape of a signal that
spends a sufficiently long time (say several months) in the eLISA band
depends quite sensitively on its intrinsic parameters (masses, spins
and eccentricity), and the motion of eLISA with respect to the source
induces amplitude and phase modulations in the signal that depend on
the position of the source and the orientation of its orbital angular
momentum. By fitting the data with precomputed templates that depend
on the different parameters of the signal one can resolve all the
source parameters. Significant correlations between the luminosity
distance and other parameters (most notably the orbital inclination)
corrupt the accuracy with which we can estimate the distance. Even so,
due to the large signal-to-noise ratios expected from supermassive BH
binaries, eLISA should be able to measure distances at $z\sim 1$ to
better than 0.1\%-1\% accuracy
\cite{holz05,Arun07b,2009CQGra..26i4027A}.

Schutz also pointed out that although GW observations can measure
distances, they are not able to measure the source redshift. Recent
work, however, has shown that at least in the case of binary neutron
stars, and quite possibly also for neutron star-BH binaries, it should
be possible to also measure the source's redshift from the effect of
tides on the waveform phasing
\cite{Messenger:2011gi,Messenger:2013}. The intrinsic mass $M_{\rm
  int}$ of the neutron star will be imprinted in the tidal effects,
and once the intrinsic mass is known the redshift can be inferred from
the fact that the observed mass $M_{\rm obs}$ is given by $M_{\rm
  obs}=(1+z) M_{\rm int}.$

For MBHBs, however, there is no way to infer the source's redshift
from GW observations alone.  Since eLISA will not be sensitive to
merging neutron star binaries (at least not at cosmological
distances), we have to rely on EM follow-up of GW events to infer the
source's redshift. Thus, it was argued that there is great synergy in
multi-messenger observations of MBH binaries (cf. Section
\ref{sec:brian}). GW observations would provide distance measurements,
while EM observations would provide redshift, and the two observations
together would be a new tool for cosmology.
\begin{figure}
            \includegraphics[width=20pc]{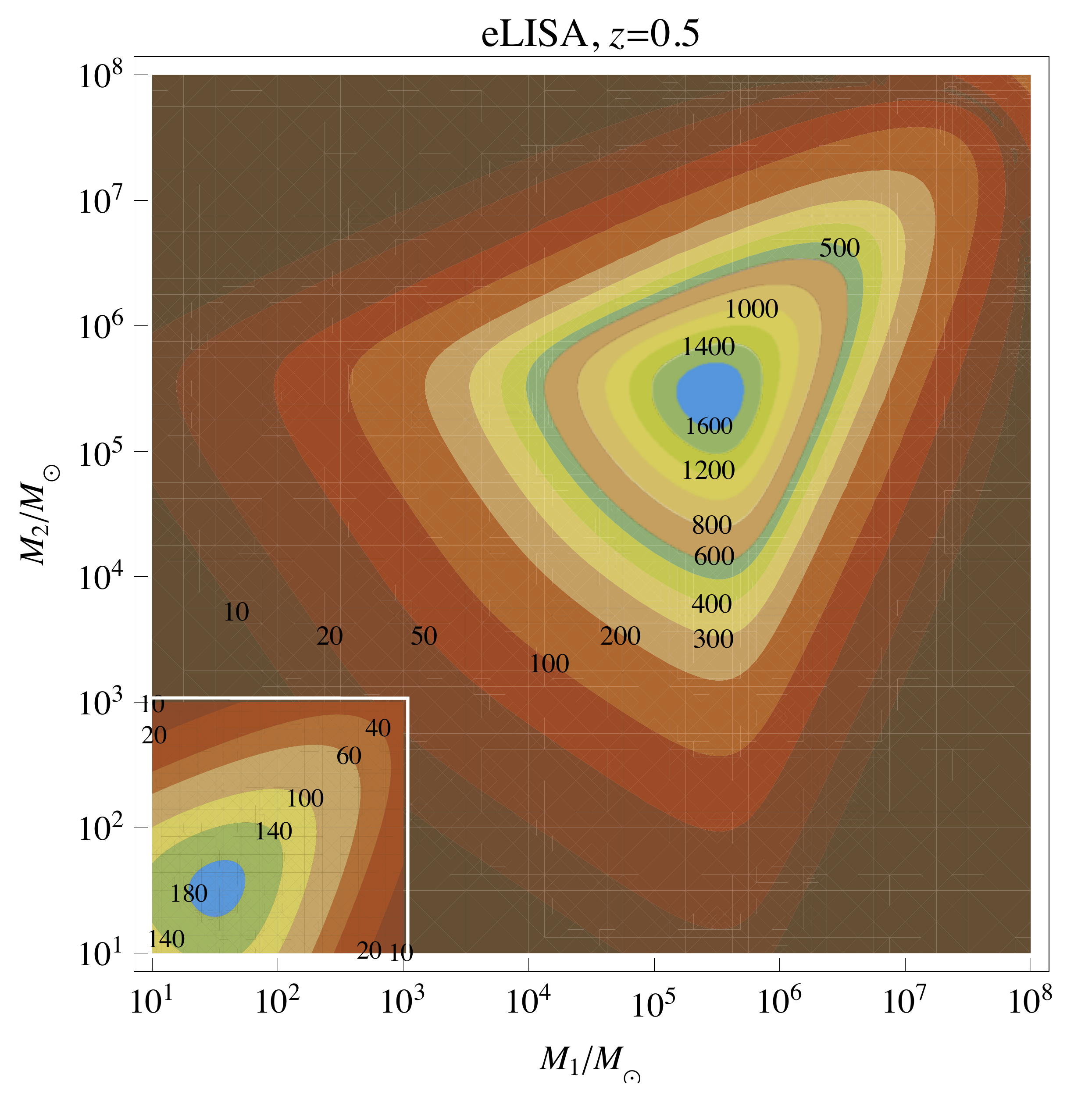}\hspace{3pc}
        \begin{minipage}[b]{15pc}
            \caption{Signal-to-noise ratio (SNR) of the inspiral phase
              of coalescing BH binaries as a function of the component
              masses of binary sources at $z=0.5$ Signals are assumed
              to last for a year before they merge in the eLISA
              band. The large SNR helps in measuring the distance with
              great accuracy in spite of correlations between other
              signal parameters. The inset shows the SNR for Einstein
              Telescope (ET)---a third generation underground GW
              detector. Although eLISA and ET are sensitive to
              different range of masses, there are some sources that
              could be observed by both eLISA and ET. }
            \label{fig:eLISA SNRs}
        \end{minipage}\hspace{2pc}%
\end{figure}

\subsection {Challenges in using eLISA for cosmography}
\label{sec:cosmography:challenges}
At first it was thought that eLISA will be a powerful new tool for
cosmography. However, it soon became clear that significant challenges
are posed by different aspects of the measurement, making the problem
quite hard \cite{holz05,VanDenBroeck:2010fp}. In the following we will
discuss the most important challenges and what solutions, if any, have
been found in confronting them.

\paragraph{Localizing sources.} 
GW interferometers like eLISA are quadrupole antennas with very good
sky coverage \cite{Sathyaprakash:2009xs}. They are able to observe
better than $2\pi$ steradians of the sky at any one time, but they are
generally not very good at resolving the sky position of the
source. The angular resolution of eLISA for sources that use only the
dominant quadrupole harmonic of the signal at twice the orbital
frequency is $\sim$ 10's of square degrees \cite{holz05}. At a
redshift of $z \sim 0.5$ (the relevant redshift where one can expect
to detect one or two merging MBHBs each year) a sky patch of that size
contains thousands of galaxies, and therefore there seems to be no
hope of identifying the host galaxy to measure the source's
redshift.
Moore and Hellings \cite{MooreHellings02,HellingsMoore03} first
realized that if one includes in the search templates higher-order
signal harmonics, in addition to the dominant harmonic at twice the
orbital frequency, the angular resolution can get sensibly smaller.
More complete analyses including higher harmonics \cite{Arun07b}
and/or spin-induced precessional effects \cite{2006PhRvD..74l2001L}
showed that sources are resolved to $\sim$ 1 square degree.
This angular resolution is much smaller than before -- not small
enough to contain just the host galaxy, but perhaps small enough to
contain the (not yet fully virialized) galaxy cluster to which the
host belongs. Moreover, higher harmonics also help break the
degeneracy between the inclination angle and the distance, reducing
the fractional error on distance measurement by a factor of 2 to 5
\cite{Arun07b}.

The Task Force that was appointed to systematically explore how well
LISA might determine source parameters and constrain dark energy
concluded that each year LISA could observe a few events for which
distance can be measured to within 1\% {\em and} the source can be
localized to within one square degree -- accuracies that are
sufficient to pin down the dark energy equation-of-state parameter $w$
at the level of one percent \cite{2009CQGra..26i4027A}. Although the
improvement in sky resolution brought through higher-order harmonics
and spin effects is sufficient to measure the host redshift, it would
be far better if sources could be identified by the EM counterpart
that the merger event might produce, which we will discuss next.

\paragraph{Electromagnetic counterparts.} 
Quite a lot of work has gone on in understanding the EM counterparts
produced by a merger event. A MBHB merger by itself does not produce
any EM radiation. The environment in which a MBHB merges consists of
hot diffuse gas, circumbinary accretion disks and stars. As
highlighted in Section \ref{sec:brian}, cold gas forming a putative
circumbinary disk can stream toward the two MBHs and form a dense,
small disk surviving at the moment of the MBHB coalescence. Numerical
simulations of MBHBs seem to suggest that the merger event could shock
and heat the surrounding gas particles to high temperatures, leading
to bright EM counterparts in coincidence to the coalescence. If the
immediate vicinity of the binary is devoided of gas at coalescence,
this sort of radiation can be expected within months to years of
merger, hence monitoring the sky position of merger could indeed be
interesting from an astrophysical point of view as well as for
cosmography (see \cite{Schnittman:2010wy} for a comprehensive review
of EM counterparts to coalescing MBHBs).

\paragraph{Weak gravitational lensing.} 
The space between a GW source and eLISA is not empty. Small-scale
inhomogeneities that are ever present in the path of a signal from a
source to eLISA render the signal brighter or dimmer, depending on the
integrated effect of the inhomogeneities along a given path. This is
called weak gravitational lensing, and all cosmological sources are
subject to this phenomenon. Weak lensing cannot be detected
accurately, and there is no way to fully correct for its effect on
particular sources.  Although weak lensing can significantly affect
individual detections, the effect averages out when considering a
large population of sources.  At redshift $z\sim 1$, systematic errors
in the measurement of distance due to weak lensing are typically $\sim
2$-$5\%$, an order of magnitude larger than statistical errors for an
average signal at this distance \cite{VanDenBroeck:2010fp}.

It is possible to correct for the systematics to some extent, but the
best possible technique (in fact a combination of different
observations) can decrease the error by a factor of 2
\cite{Shapiro:2009sr,Hirata:2010ba}.  Hence weak lensing will severely
limit how well distances can be inferred, and it is not possible to
make a meaningful measurement of dark energy with a single source. If
eLISA manages to accumulate $\gsim 30$ events at $z<2$ during its
lifetime (which implies MBHB merger rates at the upper end of what is
currently estimated and/or a mission lifetime $\gsim 5$ years) then it
should be possible to measure $w$ to an accuracy of $\sim 4$\%
\cite{Petiteau:2011we}.

\subsection{Outlook}
\label{sec:cosmography:outlook}
The foregoing summary is largely based on studies that were conducted
in the context of LISA. Future studies should repeat these
investigations in the context of eLISA to have a concrete evaluation
of its science potential. It would be useful to study what
optimizations can be made within the existing mission framework to
maximize the science potential. Estimates of event rates in the
context of eLISA, which has significantly better sensitivity than LISA
at higher frequencies (at the expense of poorer sensitivity at lower
frequencies) would be useful, as higher rates are obviously the
cleanest way to defeat the effect of weak lensing. Studies on how weak
lensing biases can be further reduced would be useful, as also more
reliable estimates of the EM radiation that would be produced after
merger. In particular, it is important to study the extent to which
gas and disks could affect the dynamics of the binary, and whether
this could bias the estimation of signal parameters.

Two concepts beyond eLISA have been studied: Big Bang Observer
\cite{Harry:2006fi} in the US and Deci-Hertz Gravitational Wave
Observatory (DECIGO) \cite{Kawamura:2011zz} in Japan. These ambitious
projects with more than an order of magnitude improvement in strain
sensitivity will obviously not have the limitations of LISA or eLISA
for cosmography (see
e.g. \cite{CutlerHolz09,2010PhRvD..81l4046H}). The problem here might
be one of data analysis. The large number of sources they might detect
could pose problems with unambiguous estimation of signal and source
parameters. It is important to study whether our knowledge of the
waveforms predicted by GR is good enough to cleanly disentangle the
signals.

\section{Black hole mergers as probes of strong-field gravity}
\label{sec:emanuele}

Einstein's GR is certainly one of the most elegant and successful
physical theories, and so far it has passed all experimental tests
with flying colors \cite{2014LRR....17....4W}. However there are
theoretical and experimental reasons suggesting that the theory must
be modified at some level.
From a theoretical point of view, GR is a purely classical theory, and
power counting arguments indicate that it is not renormalizable in the
standard field-theory sense; however, one can build a renormalizable
theory by adding quadratic curvature terms -- i.e.,
high-energy/high-curvature corrections -- to the Einstein-Hilbert
action \cite{1977PhRvD..16..953S}. Furthermore, high-energy
(ultraviolet) corrections seem necessary to avoid singularities that
are inevitable in classical GR, as shown by the Hawking-Penrose
singularity theorems \cite{1970RSPSA.314..529H}.
From an observational point of view, cosmological measurements are
usually interpreted as providing evidence for dark matter and a
nonzero cosmological constant (``dark energy''). This interpretation
poses serious conceptual issues, including the cosmological constant
problem (``why is the observed value of the cosmological constant so
small in Planck units?'') and the coincidence problem (``why is the
energy density of the cosmological constant so close to the present
matter density?''). No dynamical solution of the cosmological constant
problem is possible within GR \cite{1989RvMP...61....1W}. It seems
reasonable that ultraviolet corrections to GR would inevitably
``leak'' down to cosmological scales, showing up as low-energy
(infrared) corrections.

The arguments summarized above suggest that Einstein's theory of
gravity should be modified at both low and high energies, but it is
not easy to introduce modifications to GR that respect these
requirements without facing additional problems. Einstein's theory is
the unique interacting theory of a Lorentz-invariant massless
helicity-2 particle \cite{1970GReGr...1....9D}, and therefore new
physics in the gravitational sector must introduce additional degrees
of freedom. 
Any additional degrees of freedom must modify the theory at both low
and high energies {\em while being consistent with GR in the
  intermediate-energy regime}, i.e. at length scales between $\sim
1\,\mu$m and about one Astronomical Unit, where the theory is
extremely well tested. Intermediate-energy constraints include
laboratory experiments, Solar System experiments (that verify the
Einstein Equivalence Principle to remarkable accuracy, and force
parametrized post-Newtonian parameters such as $\beta$ and $\gamma$ to
be extremely close to their GR values) and binary pulsar experiments
(that place stringent bounds on popular extensions of GR such as
scalar-tensor theories, Lorentz-violating theories and TeVeS): see
\cite{2014LRR....17....4W,2014arXiv1402.5594W} for reviews.

eLISA is arguably the best strong-gravity laboratory one could hope
for, as it will probe the strong-field dynamics of GR out to
cosmological distances to levels unachievable by binary
pulsars\footnote{One of the most extraordinary laboratories for
  strong-gravity tests is PSR J0348+0432
  \cite{2013Sci...340..448A}. Even this binary system, which is highly
  relativistic for binary-pulsar standards, has an orbital velocity
  $v\simeq 2\times 10^{-3} c$, much smaller than the orbital
  velocities $v\approx c/3$ of an astrophysical BH binary near
  merger.}, Earth-based interferometers or Pulsar Timing Arrays (see
\cite{2013LRR....16....7G,2013LRR....16....9Y} for reviews). As we
will argue below, space-based GW interferometers could
provide crucial hints about the nature of the compact objects at
galactic centers and about the nature of gravity itself.

It is useful to classify ``tests of strong field gravity'' as
belonging to two -- qualitatively very different -- categories:

\noindent
{\em (1) External tests: can laboratory experiments, astrophysical
observations or future GW measurements determine whether GR is the
correct theory of gravity?} To frame this question in terms of
hypothesis testing, one would like to have a valid opponent to
GR. What constitutes a ``valid opponent'' is a matter of taste. For
our purpose (i.e., tests of strong-field gravity with BH
mergers) it should be a cosmologically viable fundamental theory with
a well-posed initial value formulation, and field equations that
follow from an action principle. Furthermore, the theory should be
simple enough to allow calculations of (say) BH solutions and
GW emission.  There are countless attempts to modify GR
\cite{2012PhR...513....1C}, but all modifications must introduce some
sort of screening mechanism in order to be viable at intermediate
energies. Screening mechanisms include chameleons, symmetrons,
dilatons,
MOND-like dynamics,
the Vainshtein mechanism, etcetera
\cite{2014arXiv1407.0059J}. Since we don't have a full theory of
quantum gravity, an effective field-theory approach is often invoked
when constructing phenomenological alternatives to GR
\cite{2004LRR.....7....5B,2007ARNPS..57..329B}. For example, one can
start with the most generic four-dimensional theories of gravity
including quadratic curvature invariants generically coupled to a
single scalar field $\phi$:
\begin{align}
    S&=\int\frac{d^4x\,\sqrt{-g}}{16\pi} \left[R-2\nabla_a\phi\nabla^a\phi-V(\phi)+f_1(\phi)R^2
+f_2(\phi) R_{ab}R^{ab}+f_3(\phi) R_{abcd}R^{abcd}+f_4(\phi){}^{*}\!RR\right]\nn\\
&+S_\text{\rm mat}\left[\gamma(\phi)g_{\mu\nu},\,\Psi_\text{\rm mat}\right]\,,\label{action_quadratic}
\end{align}
where $\Psi_\text{\rm mat}$ collectively denotes matter fields,
$V(\phi)$ is the scalar self-potential, $f_i(\phi)$ are generic
coupling functions, the Chern-Simons term
${}^{*}\!RR\equiv\frac{1}{2}R_{abcd}\epsilon^{baef}R^{cd}_{~~ef}$,
$\epsilon^{abcd}$ is the Levi-Civita tensor, and in the matter action
$S_\text{mat}$ we allow for a nonminimal coupling that violates the
(weak) equivalence principle. Not all of these theories are
acceptable: for example, to avoid higher-order derivatives in the
equations of motion one must generally assume the couplings to be
small and treat the theory as an effective field theory (the equations
are second-order in the strong-coupling limit only if the quadratic
invariants enter in the special ``Gauss-Bonnet'' combination). The
action above may seem complicated, but it actually represents a very
restricted class of theories, and calculations of gravitational
radiation have been performed only in very specific subcases, such as
scalar-tensor theories with specific choices of the potential and
couplings
\cite{1996PhRvD..54.1474D,2012PhRvD..85f4041A,2013PhRvD..87h4070M,2014PhRvD..89h4014L}
and some forms of quadratic gravity
\cite{2012PhRvD..85f4022Y,2013PhRvD..87h4058Y}.  The bottom line is
that there are very few ``serious'' alternatives to GR (in the sense
that they are well posed, follow from a Lagrangian, make sensible
predictions...) and even fewer for which GW calculations have been
carried out. For these theories, GW observations usually yield
constraints that are comparable to, and often better than, binary
pulsar and Solar System bounds
\cite{2013LRR....16....7G,2013LRR....16....9Y}.

\noindent
{\em (2) Internal tests: is GR ``internally'' consistent with
  astrophysical observations?} One of the most striking predictions of
GR is the existence of BHs. Astronomers commonly believe that the
compact objects that harbor galactic centers are the BHs of GR, but
this ``BH paradigm'' rests on somewhat shaky foundations. Evidence
that these objects possess event horizons (or more correctly, apparent
horizons) rather than solid surfaces usually rests on plausibility
arguments based on accretion physics
\cite{2005NJPh....7..199N,2013arXiv1312.6698N}, that leave room for
some skepticism\footnote{From a theorist's point of view, one of the
  most convincing arguments in favor of the BH paradigm is that the
  alternatives are either unstable (as in the case of dense star
  clusters, fermion stars or naked singularities), unnatural
  (e.g. ``exotic'' matter violating some of the energy conditions),
  contrived (such as gravastars), implausible as the end-point of
  collapse in astrophysical settings (boson stars) or nearly
  indistinguishable from Kerr (this is the case for BH solutions in
  alternative theories with coupling parameters that are reasonable
  from a fundamental physics point of view).}
\cite{2002A&A...396L..31A}.
It is also important to stress that, strictly speaking, any tests that
probe the Kerr {\em metric} alone (such as tests based on matter
accretion or ray-tracing of photon trajectories) are of little value
as internal tests of GR. The reason is that most alternative theories
(including generic scalar-tensor theories \cite{2012PhRvL.108h1103S}
and a large class of higher-curvature theories
\cite{2008PhRvL.100i1101P}) admit the Kerr metric as a solution, and
the theories that don't (e.g. Einstein-dilaton Gauss-Bonnet
\cite{2011PhRvL.106o1104K}, Dynamical Chern-Simons
\cite{2014arXiv1405.2133A}, and Lorentz-violating gravity~\cite{2011PhRvD..83l4043B,2012PhRvL.109r1101B,2013PhRvD..87h7504B,2013CQGra..30x4010B}) predict BH solutions that differ from GR
by amounts that should be astrophysically unmeasurable (see, however, Refs.~\cite{2014PhRvL.112v1101H,2014PhRvD..90f5019D} for a family of BH solutions whose
deviations from the Kerr metric may be important). Many
``quasi-Kerr metrics'' that have been proposed in this context should
be viewed as unnatural strawmen: they often have serious pathologies
\cite{2013PhRvD..87l4017J}, and they are therefore unacceptable even
for the limited scope of parametrizing deviations from the Kerr metric
\cite{2014PhRvD..89f4007C}.

These considerations imply that the only way to unambiguously verify
that the compact objects in galactic centers are actually Kerr BHs is
{\bf via their GW dynamics}, especially in the strong-field
merger/ringdown phase
\cite{2006PhRvD..73f4030B,2007PhRvD..76j4044B,2008PhRvL.101i9001B}. This
is why space-based GW observations hold great promise to constrain
strong-field gravity \cite{2013LRR....16....7G}. Their qualitative
advantage over Earth-based detectors is simple to understand. The
fundamental oscillation mode of a nonrotating BH has frequency
$f\simeq 1.2\times 10^{-2} (10^6M_\odot/M)$~Hz. This frequency lies
exactly in the ``bucket'' of eLISA's noise power spectral density for
``light'' BHs of mass $M\sim 10^6M_\odot$, that were presumably the
building blocks of the large BHs we see at galactic centers
today. Advanced LIGO, by contrast, has maximum sensitivity at $f\sim
10^2$~Hz, i.e. for intermediate-mass BHs of mass $M\sim 10^2M_\odot$,
whose very existence is still highly uncertain
\cite{2004IJMPD..13....1M}.
These oscillation modes are called ``quasinormal'' modes, because they
are damped by GW emission. The no-hair theorem stated in Section
\ref{sec:enrico} implies that, in GR, the frequencies and damping
times of all QNMs depend only on the BH mass $M$ and spin $a$. A
measurement of the dominant mode's frequency and damping time yields
both $M$ and $a$; the measurement of {\em any} other frequency and/or
damping time can then be used to verify that the BH formed as a result
of the merger is indeed a Kerr BH, as predicted by GR. The feasibility
of this (internal) test of GR depends on the measurability of QNM
frequencies/damping times and on our ability to resolve modes. Both
measurability and resolvability scale like $1/\rho$, where $\rho\sim
h/S_n$ is the signal-to-noise ratio of the merger event
\cite{2006PhRvD..73f4030B,2007PhRvD..76j4044B}. The maximum
sensitivity $S_n$ of Earth-based and space-based detectors is
comparable in order of magnitude, but (as we saw above) space-based
detectors target sources that are $\sim 10^4$ times more massive. The
GW amplitude $h\sim \sqrt{\epsilon_{\rm rd}M}$, where $\epsilon_{\rm
  rd}\sim 10^{-2}(4\eta)^2$ \cite{2007PhRvD..76f4034B} is a ``ringdown
efficiency'' and $\eta\equiv m_1 m_2/(m_1+m_2)^2\in [0,\,0.25]$ is the
so-called symmetric mass ratio. The punchline of this argument is that
$\rho\sim \sqrt{M}$, so {\em no-hair theorem tests with space-based
  observations of BH binary mergers are typically $\sim 10^2$ stronger
  than Earth-based GW tests}.

In summary, BH mergers are extraordinary (local) probes of the no-hair
theorem that are potentially detectable by eLISA throughout the entire
Universe. Each merger event allows us to do much more than this: it
can give us tests of consistency of the post-merger Kerr remnant with
the pre-merger binary dynamics
\cite{2012PhRvD..85b4018K,2012PhRvD..85l4056G,2012PhRvL.109n1102K},
allow us to test the area theorem \cite{2005ApJ...623..689H}, and
perhaps even allow us to peer into the nonlinear dynamics responsible
for the coupling of different quasinormal modes
\cite{2014arXiv1404.3197L}.
Even more interestingly, each BH binary merger can be thought of as a
local probe of whether {\em strong-field} GR is valid {\em at the
  redshift $z$ at which the merger occurred}. This is an incredible
opportunity for tests of GR, because it would verify that Einstein's
gravity accurately describes BHs at least out to redshift $z$. This
would place even more stringent constraints on the viable
modifications of GR. This idea has one drawback: it requires (ideally)
the determination of the merging binary's redshift $z$, or at the very
least the determination of a {\em lower bound} on the source
redshift. As described in Section \ref{sec:sathya}, source
localization and distance determination are intimately related in a
LISA-like mission, and they get significantly better for a three-arm
mission \cite{2009CQGra..26i4027A}. In conclusion, a three-arm mission
would make a big difference to test the no-hair theorem at
cosmological redshift - and thus to place tight constraints on the
strong-field behavior of the theory at distances where cosmological
observations are relevant.

Another aspect worth emphasizing is that, unlike Solar System tests
(that only probe the ``static'', quasi-Newtonian behavior of
gravitational fields) and binary pulsars (that essentially measure the
energy flux predicted by a given theory of gravity, and not much else
in terms of the dynamics of the gravitational field) the {\em direct}
observation of GWs can probe the number of polarization states as well
as the propagation properties of GWs, as encoded in their dispersion
relation.  Tests of the dispersion relation place constraints on a
putative nonzero graviton mass, and they get better, as expected, for
sources at cosmological distance. In fact it has been shown that eLISA
can constrain the mass of the graviton to a level that is $\sim 4$
orders of magnitude better than current Solar System constraints
\cite{1998PhRvD..57.2061W,2005PhRvD..71h4025B}. Quite naturally,
constraints on any given alternative theory of gravity get better with
multiple observations because of the improved statistics. Calculations
in the case of a hypothetical massive graviton show that the
improvement is better than the naive Poisson-statistic expectation of
$\sqrt{N}$, where $N$ is the number of events, because the louder
events ``carry more weight'' in determining the combined bound
\cite{2011PhRvD..84j1501B}.

Last but not least, the discovery space of eLISA is potentially
enormous. The potential of eLISA for cosmology was discussed in
Section~\ref{sec:sathya}; here we will focus on the discovery space
related to MBH observations. Binary pulsars are already constraining
phenomena like ``spontaneous scalarization'' to levels that are
comparable to what eLISA could do by observing neutron stars spiraling
into MBHs \cite{2005PhRvD..71h4025B}. However, one can imagine
scenarios where modified gravity would produce smoking-gun signatures
of deviations from GR that would be observable by eLISA, and
completely invisible to binary-pulsar tests. The simplest case study
are extreme mass-ratio inspirals in scalar-tensor theory. If the
scalar has a mass, Kerr BHs become vulnerable to the so-called
``black-hole bomb'' instability: superradiance can amplify incident
waves, the mass of the scalar acts like a mirror that reflects
amplified waves back onto the BH, and therefore the superresonant
amplification of incident waves can grow without bound
\cite{1972Natur.238..211P}. This instability could have striking
effects, such as the existence of ``floating orbits''
\cite{2011PhRvL.107x1101C} at which the inspiralling body could stall,
emitting essentially monochromatic radiation -- a ``GW laser''!
Similar effects could occur if the energy conditions are violated
\cite{2013PhRvL.111k1101C,2013PhRvD..88d4056C}, so eLISA could hint at
exotic physics responsible for possible violations of the energy
conditions. Finally, if the objects at galactic centers were not BHs
but (say) gravastars or boson stars, the oscillation modes of these
exotic objects would inevitably be excited by orbiting bodies, leaving
characteristic signatures in the energy flux that are potentially
detectable by eLISA \cite{2010PhRvD..81h4011P,2013PhRvD..88f4046M}.

Opening new observational windows on the Universe inevitably reveals
more than we anticipated. We should be ready for surprises.

\section{Conclusions}

We have reviewed our current understanding of MBH formation and
evolution. The mysteries that surround the birth and growth of these
cosmic monsters are enormous, and they are intimately related to the
growth of structure in our Universe. Observations of gravitational
radiation from MBH mergers -- especially in combination with EM
counterparts -- have the potential to measure spins to levels
unachievable by other means, clarify the role of accretion in MBH
growth, and constrain fundamental physics in unprecedented ways. An
eLISA-like mission will be a spectacular, unrivalled laboratory for
fundamental physics and astrophysics.

\ack Enrico Barausse and Alberto Sesana thank their co-authors Massimo
Dotti and Elena Maria Rossi for insightful conversations.  Enrico
Barausse acknowledges support from the European Union's Seventh
Framework Programme (FP7/PEOPLE-2011-CIG) through the Marie Curie
Career Integration Grant GALFORMBHS PCIG11-GA-2012-321608.  Emanuele
Berti is supported by NSF CAREER Grant No. PHY-1055103.  Bangalore
S.~Sathyaprakash acknowledges the support of the LIGO Visitor Program
through the National Science Foundation award PHY-0757058 and STFC
grant ST/J000345/1.

\vspace{.5cm}

%
%


\section*{References}


\providecommand{\newblock}{}

\end{document}